\documentclass[runningheads]{llncs}
\usepackage{graphicx}
\usepackage{subcaption,ragged2e}
\usepackage[ruled]{algorithm2e}

\usepackage{comment}
\usepackage{float}

\begin{document}
\title{Temporal and Geographical Analysis of Real Economic Activities in the Bitcoin Blockchain}
\titlerunning{Bitcoin Temporal and Spatial Analysis}
%
\author{Rafael Ramos Tubino\inst{1} \and
Rémy Cazabet\inst{1} \and
Natkamon Tovanich\inst{2} \and \\
Céline Robardet\inst{1}}
\authorrunning{R. Ramos Tubino et al.}
%
\institute{Univ Lyon, UCBL, CNRS, INSA Lyon, LIRIS, UMR5205, \\F-69622 Villeurbanne,
France \and
Blockchain@X, CREST, École Polytechnique,
Institut Polytechnique de Paris, \\F-91120 Palaiseau, France
}
\maketitle              
\begin{abstract}
We study the \textit{real economic activity} in the Bitcoin blockchain that involves transactions from/to retail users rather than between organizations such as marketplaces, exchanges, or other services. We first introduce a heuristic method to classify Bitcoin players into three main categories: Frequent Receivers (FR), Neighbors of FR, and Others. We show that most real transactions involve Frequent Receivers, representing a small fraction of the total value exchanged according to the blockchain, but a significant fraction of all payments, raising concerns about the centralization of the Bitcoin ecosystem. We also conduct a weekly pattern analysis of activity, providing insights into the geographical location of Bitcoin users and allowing us to quantify the bias of a well-known dataset for actor identification.

\end{abstract}
\section{Introduction}
Bitcoin is a public cryptocurrency blockchain introduced in 2009. 
It provides a rich, publicly accessible repository of cryptocurrency transactions that is a subject of interest both in academia and the general public.
While numerous new usages of blockchains have developed in recent years, e.g., Decentralized Finance (DeFi), Non-fungible Tokens (NFTs), smart contracts, their usage as a currency, or at least as a store and exchange of value, remains one of the most important.
As of June 2023, Bitcoin transactions still dominate financial cryptocurrency transactions, solidifying its position as the crypto asset with the highest market capitalization and daily transaction volume.

Understanding the nature of transactions occurring in a cryptocurrency such as Bitcoin is an important question. Since Bitcoin relies on a public blockchain, anyone can access and collect data on all transactions between Bitcoin users. However, due to Bitcoin pseudonymity 
and to the presence of transactions of different purposes,
it is not trivial to interpret the nature of transactions taking place. In this work, we propose to analyze transactions that could be assimilated to “retail” activities or, more broadly, to activities triggered by non-professional Bitcoin players, being to pay other users, a company, or involving an exchange platform. Our objective is to monitor the usage of Bitcoin compatible with being a \textit{currency}, a \textit{real economy}, as opposed to financial trading activities, technical transactions, and transactions not involving an actual exchange of value with at least one non-professional Bitcoin user.

\section{Related Works}
Bitcoin transactions have been studied for more than a decade.
A seminal work~\cite{meiklejohn2013fistful} analyzed Bitcoin activity by trying to evaluate the fraction of transactions made by exchange or gambling platforms. However, this work, carried out in the early years of Bitcoin, required the manual labeling of transactions and thus involved only a handful of well-known users.
More recent approaches with a similar goal~\cite{liu2021characterizing} applied machine learning techniques to automatically classify users into categories based on their transaction pattern and then use these categories to analyze the type of ongoing activities in the Bitcoin economy. A weakness of these approaches is that they rely on small labeled datasets (mostly WalletExplorer~\cite{walletexplorer}) to be able to train machine learning methods. 
Other works focus on some particular types of activities or users, such as ransomware~ \cite{dalal2021identifying}, dark marketplaces~\cite{hiramoto2020measuring} or money laundering~\cite{lorenz2020machine}.
A large body of research focuses on the overall quantitative analysis of the transaction network and seeks to assess the distribution of wealth~\cite{kondor2014rich}, understand the organization of the network, or measure its growth~\cite{tao2021complex,di2018data}.
However, these works did not distinguish between the variety of players and types of transactions, i.e., a large money laundering or financial trade transaction is considered the same as a small transaction between users.
To analyze individual Bitcoin players’ behaviors in a more reliable way, some authors instead rely on field surveys \cite{ante2022profiling,ante2022individual,nemeczek2023insights}. 
However, this type of approach does not provide a global view of the Bitcoin ecosystem and is limited to a handful of participants, which are difficult to identify and reach without biasing the results.
The closest work to our paper~\cite{aoyama2022cryptoasset} examines the behavior of regular users who frequently appear on a weekly basis within a specific period of interest. However, its primary focus is limited to studying the occurrence of bubbles and the participation of these users during certain weeks.

\section{Materials and Entity Types Definition}
Since Bitcoin blockchain data is publicly available, we downloaded the full blockchain data by installing a Bitcoin node, extracting transaction details using bitcoin-etl\footnote{\url{https://github.com/blockchain-etl/bitcoin-etl}}, and creating a database suitable for our needs~\cite{tubino2022towards}. We enriched the data with the daily price of Bitcoin in USD.

\subsection{Identifying real economic transactions}
To analyze the real economic activity and its evolution in Bitcoin cryptocurrency, we first need to distinguish transactions corresponding to an exchange of value between two different entities (called “payments”) from artificial transactions done for other reasons. Here, we describe the various processing steps and how they are relevant to filter out artificial transactions.

\vspace{5pt}


\noindent\textbf{Address clustering.}
The blockchain identifies transactions between pseudonymous Bitcoin addresses, cryptographic public keys. We use the common-input heuristic~\cite{harrigan2016unreasonable}, that states that all inputs of a given transaction are owned by the same entity. To discover clusters of addresses, each cluster corresponding to a unique \textit{Bitcoin entity}, of unknown nature (e.g., companies, private individuals).

\vspace{3pt}

\noindent\textbf{UTXO outputs}
Bitcoin transactions do not necessarily correspond to a single payment from one entity to another, but frequently have several \textit{UTXO~\cite{nakamoto2008bitcoin} outputs}. Each of these outputs is a payment, and we first create a dataset of individual \textit{payments}, from the source entity --~unique thanks to the common-input heuristic~-- to each of the output entities. The number of unique payments is thus larger than the number of Bitcoin transactions as stored in the blockchain.

\vspace{3pt}

\noindent\textbf{Change.} Because of the UTXO~\cite{nakamoto2008bitcoin} mechanism, actors need to send back change to themselves, creating artificial coins exchange without economic signification. We remove these transactions that we are able to identify by removing \textit{self-transactions}, between the same Bitcoin entity.

\vspace{3pt}

\noindent\textbf{Dust \& Micro Outputs.} A phenomenon which has been identified and described is the use of \textit{dust}, small-amounts sent often to many recipients, for a variety of reasons, for instance \textit{forced address reuse}~\cite{loporchio2023bitcoin}. We think that, more generally, due to the high transaction fees and lack of technical solutions to use Bitcoin for micro-payments, all small amount payments can be considered as noise and discarded. Although the amounts are small, these transactions can bias our data as they are typically sent in very large numbers. We thus fix a minimum threshold of 0.5 USD, below which a transaction output is removed from our payment dataset.

\vspace{3pt}

\noindent\textbf{Macro Outputs.}
We also get rid of transactions with very large amounts, as they may represent unconventional real payments. Those large payments may correspond to cash management between addresses of the same entity, combined payments between exchanges, or any other type of transactions that are not between customers and businesses and customers and customers. We set a conservative upper limit of 10,000 USD, assuming that transactions beyond this amount can be assimilated to professional investor profiles, even if they are carried out by individual users.

\vspace{3pt}

\noindent\textbf{Clarification on Trading.}
A common belief about Bitcoin is that there is no "real" activity, most transactions being due to trading, that we would not consider in our analysis. Luckily, trading activities in Bitcoin are nearly exclusively performed by Exchange platforms (e.g., Binance or Kraken). Such transactions are conducted by internal scripture and never written to the blockchain. Many transactions that we observe are certainly linked to private trading activities, but only indirectly: customers moving their capital out of an Exchange to a privately owned Bitcoin address, and \textit{vice versa}.

\subsection{Defining entity types}
Our objective is to identify entities involved in real economic exchanges, such as Gambling services, Retail outlets, Exchanges acting as retail banks, etc. on the one hand, and "normal users" interacting with them in the other hand; a third category regrouping addresses used for other purposes (e.g., technical transactions). Despite existing works mentioned in the state of the art, our experiments make us think that it is impossible to reliably distinguish all or most of these entities based on classification tasks. Normal users because they have little interactions, and companies because of the emergence of diversified super-players such as Binance, which offer multiple services at once, and the general difficulty to recognize entities based only on transaction patterns.

We thus propose a simpler distinction between three entity types, that we think can be more reliable, and is sufficient to identify their role in genuine value exchange. The first type is simply defined by a certain level of activity, which is easy to observe in blockchain data, while the other types are defined relatively to this first one.
More formally, we define these categories as follows:\
%

\textbf{Frequent Receivers (FR).} 
    Entities who receive a steady stream of payments over a period of time, probably businesses. We define that this type of user must receive at least one genuine payment every day for 20 days in a month (which leaves two possible closing days per week). This criterion should exclude any non-professional user, who does not receive payments every day. In addition to this, a business user must have expenses related to his business activity. He must therefore participate in 10 other transactions, either as emitter or receiver.

\textbf{First Neighbors of FR (N1).} This category identifies customers of FR entities, i.e., private owners. These are entities that do not meet the previous conditions, but trade (pay or receive) with entities classified as FR. We include transactions in which an actor of type N1 \textit{receives} payments from one of FR type, because it can be a refund, a prize, in the case of gambling, or simply a transfer to himself from an Exchange or a Wallet manager.

\textbf{The Others (TO).} They are the entities that are not included in the two previous categories, they represent the deep part of Bitcoin, that we cannot identify as companies or private owners.

We have classified entities into each category on a monthly basis. Since an entity can change its behavior over time, it can be classified differently at two time periods. The reasons for a change in an entity's activity may be of different natures: entity clusters may no longer be used to receive or send payments, changes in the entity's internal policy may occur, or this change may be related to specificities of the entity's market.

\section{Results}
\subsection{Evolution of the Volume of Genuine Transactions}
\begin{figure}[htb]
\centering
\begin{tabular}{cc}

    \includegraphics[width=0.5\textwidth]{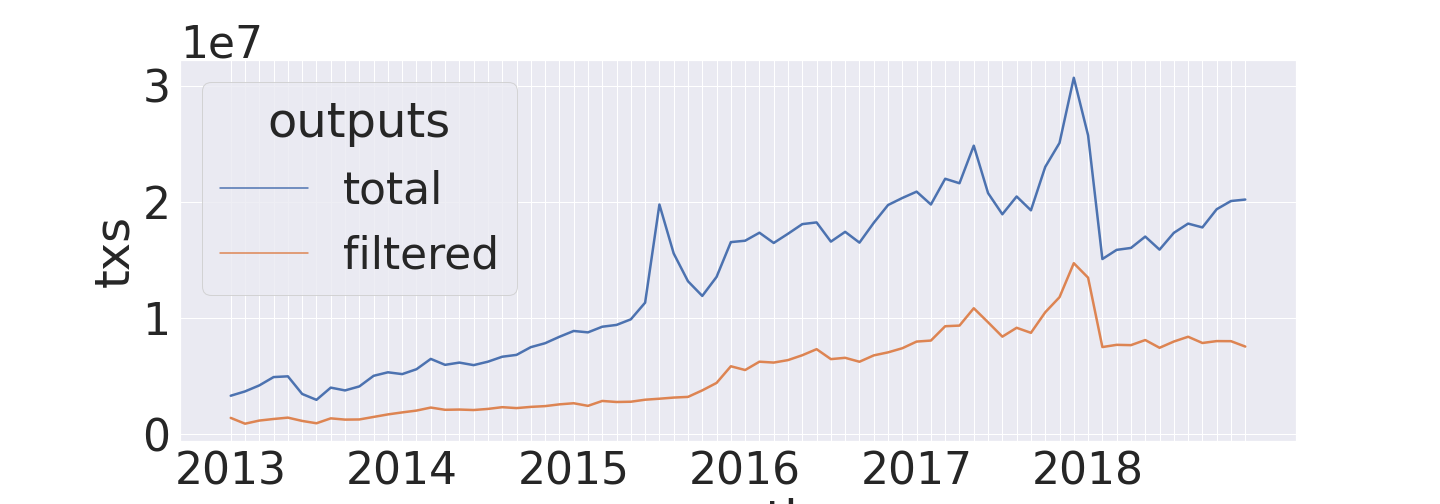}
    &\includegraphics[width=0.5\textwidth]{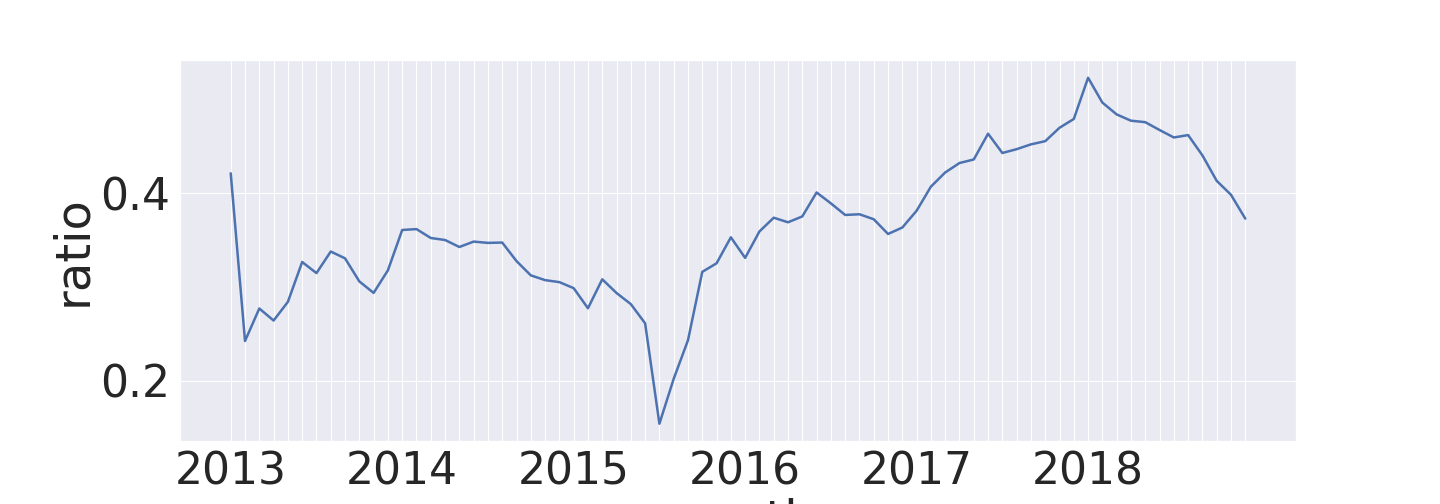}
\end{tabular}
\caption{(Left) Total payment volume. (Right) Ratio of filtered to total payments.}
\label{img_total}
\end{figure}

\noindent\textbf{Payment volume.}
We first study the relationship between the total volume of all payments in Bitcoin compared to the real (filtered) payments defined in the previous section. Figure \ref{img_total} (left) shows the evolution of the number of payments for each month in the studied period. Figure \ref{img_total} (right) outlines the difference by plotting the ratio of the two. The ratio is stable around a mean of 0.36, with maximum and minimum values of respectively 0.52 in January 2018 and 0.15 in July 2015. While both values tend to grow with time, they also undergo important differences. For instance, in July 2015, there was a peak in the total number of payments (Fig. \ref{img_total} (left)), but not in our filtered payments, resulting in the minimum value. After the growth and sharp decline in February 2018, the total volume of transactions grew while the real economy remained constant(Fig. \ref{img_total} (left)).

\begin{figure}[htb]
\centering
\captionsetup[subfigure]{justification=Centering}

\begin{subfigure}[t]{0.99\textwidth}
    \includegraphics[width=\textwidth]{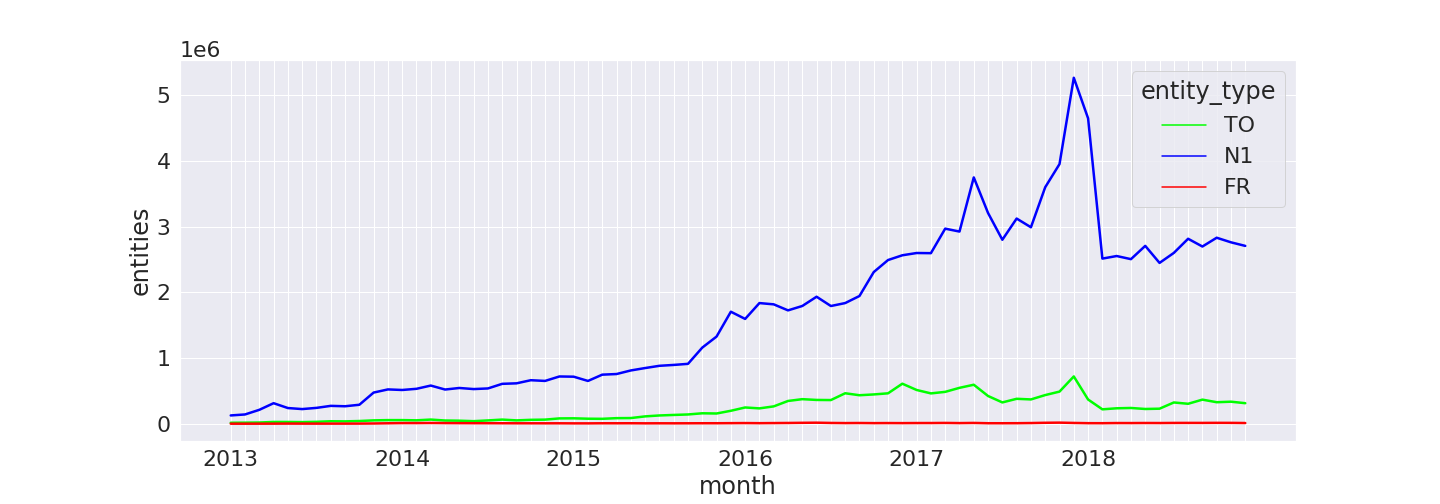}
    \caption{Number of entities in each category.}
    \label{img_cats_nbs}
\end{subfigure}
\begin{subfigure}[t]{0.49\textwidth}
    \includegraphics[width=\textwidth]{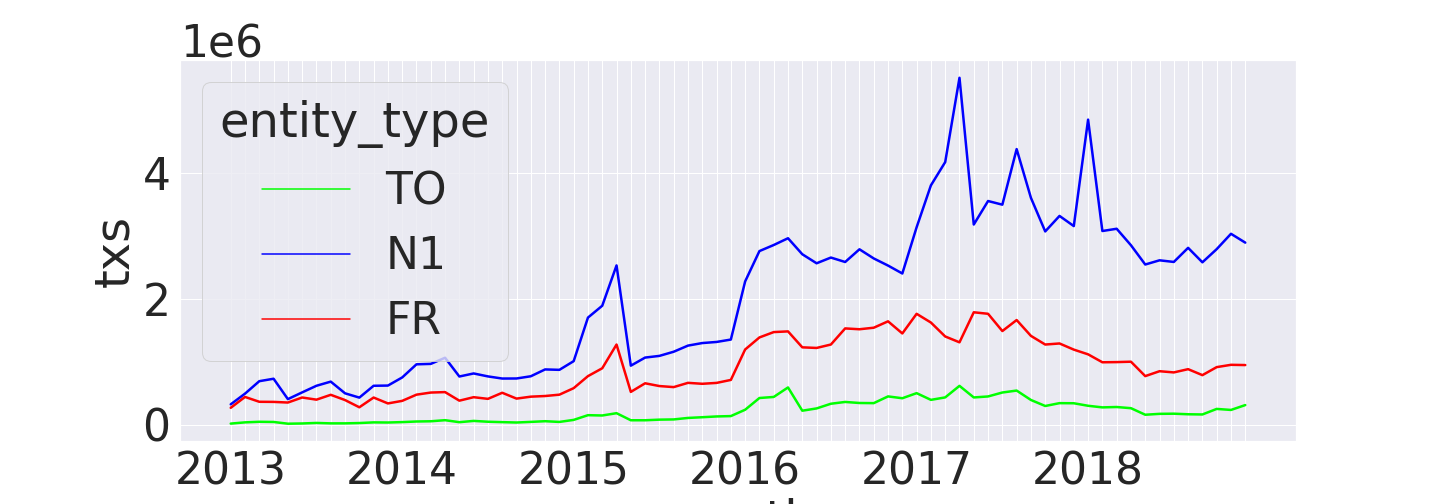}
    \caption{Sum of payment sent (input).}
    \label{img_cats_ins}
\end{subfigure}
\begin{subfigure}[t]{0.49\textwidth}
    \includegraphics[width=\textwidth]{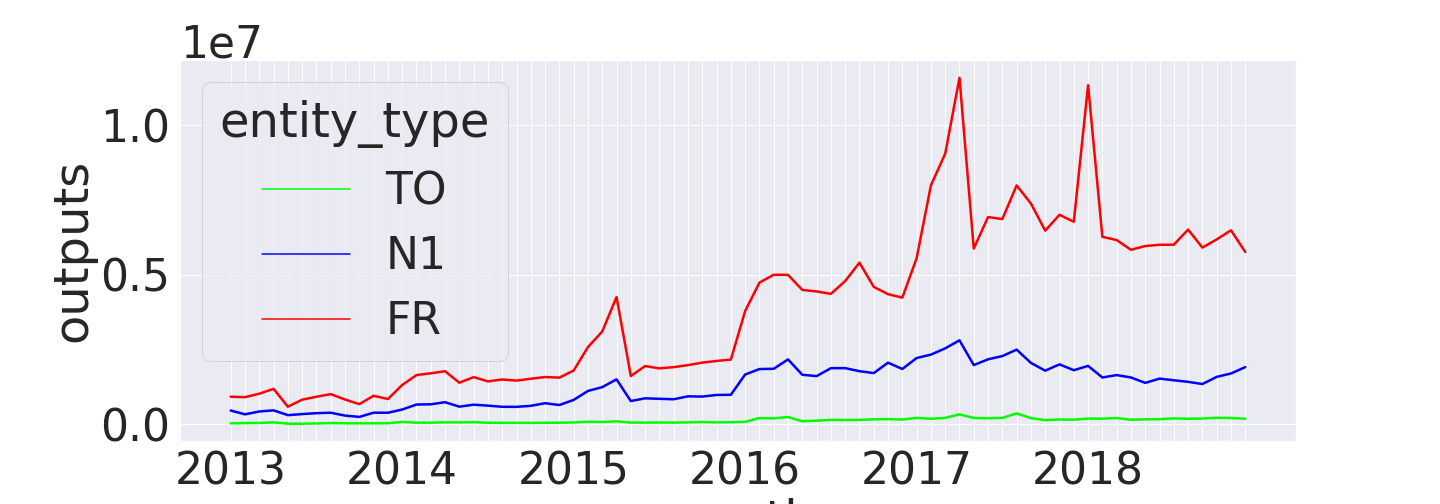}
    \caption{Sum of payment received (output).}
    \label{img_cats_outs}
\end{subfigure}
\begin{subfigure}[t]{0.49\textwidth}
    \includegraphics[width=\textwidth]{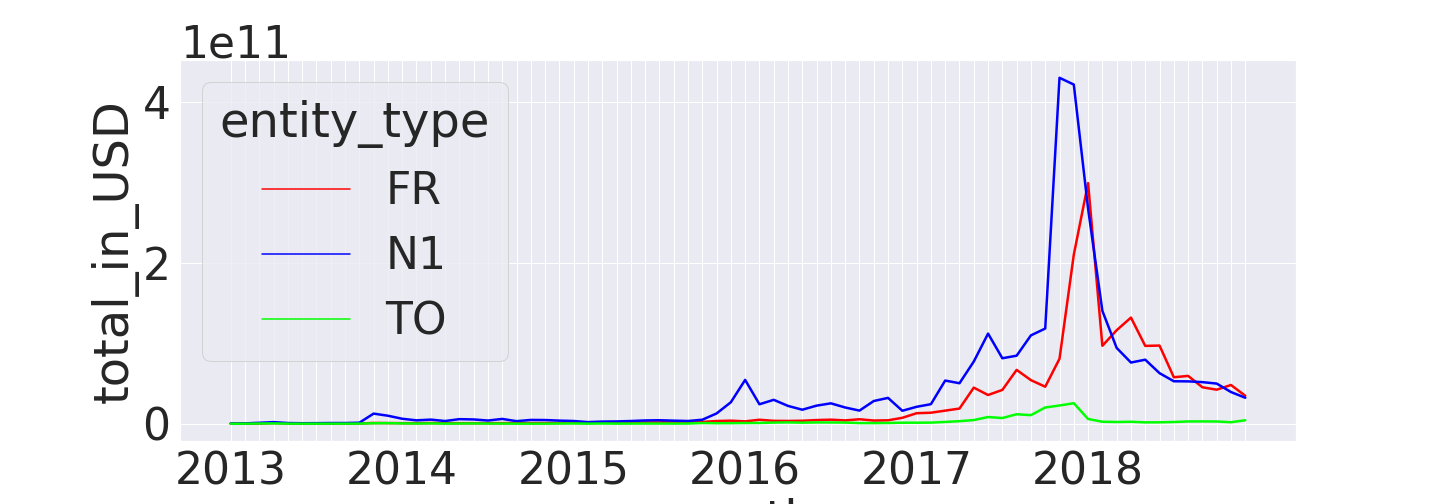}
    \caption{Sum of USD sent for each category.}
    \label{img_cats_src_usd}
\end{subfigure}
\begin{subfigure}[t]{0.49\textwidth}
    \includegraphics[width=\textwidth]{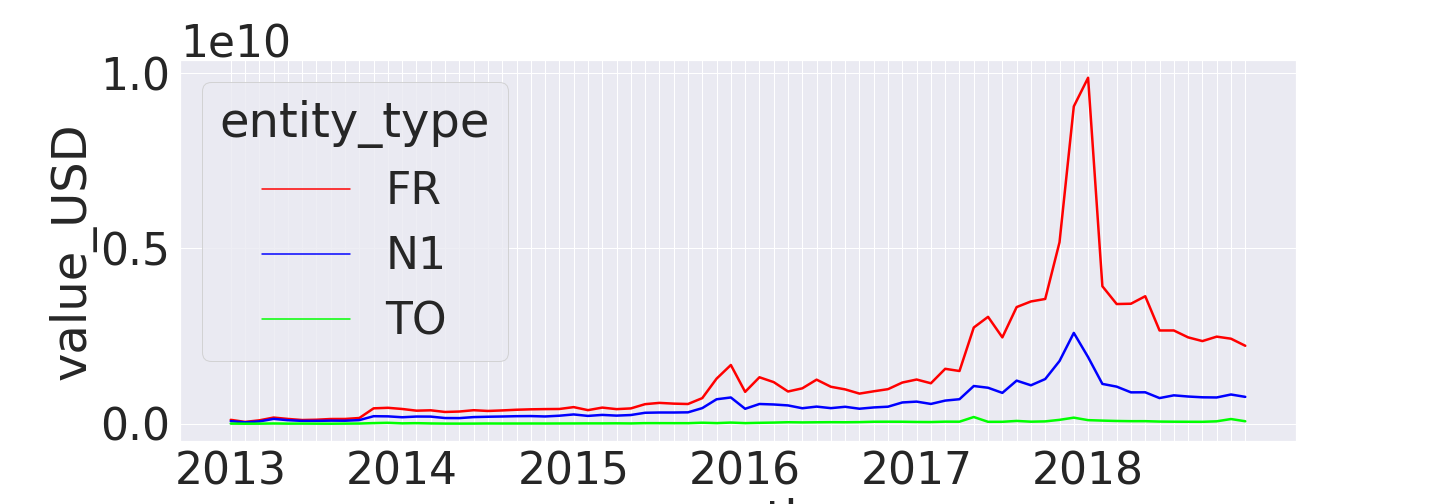}
    \caption{Sum USD received for each category.}
    \label{img_cats_dst_usd}
\end{subfigure}
\caption{Activity by category from Jan. 2013 to Dec. 2018.}
\end{figure}

\noindent\textbf{Payments entity types.}
We then analyze the prevalence of the three categories. Figures \ref{img_cats_nbs}, \ref{img_cats_ins} and \ref{img_cats_outs} show the number of entities, the volume of payment sent and received in each category, respectively.
In addition, Figures \ref{img_cats_src_usd} and \ref{img_cats_dst_usd} depict the amount of USD sent and received by category. These figures indicate that (1) FR entities are the least common but receive the most payments and the largest amounts;
(2) The N1 entity type has the highest number of entities and is the category that sends the most payments with the largest amounts (although close to FR); and
(3) TO (The Others) are the least present in transactions and are the ones who trade the smallest volume in USD.

Given the sulfurous reputation of cryptocurrency, it can be surprising that so few actual payments occur on the deep part of the Bitcoin economy(TO). Most of the real payments are received by big players (FR) and sent by entities that interact with those big players. Moreover, we found that the behavior of each category is not driven by a global tendency. For example, in Figure \ref{img_cats_ins}, we can observe that the number of payments where N1 entities appear as a sender suffers a huge rise during the first months of 2017, while we do not see the same effect for the other categories. This observation also aligns with Figure \ref{img_cats_outs} for the number of payments from FR entities.

\begin{figure}[htb]
\centering
    \includegraphics[width=0.75\textwidth,trim={0 30px 0 0},clip]{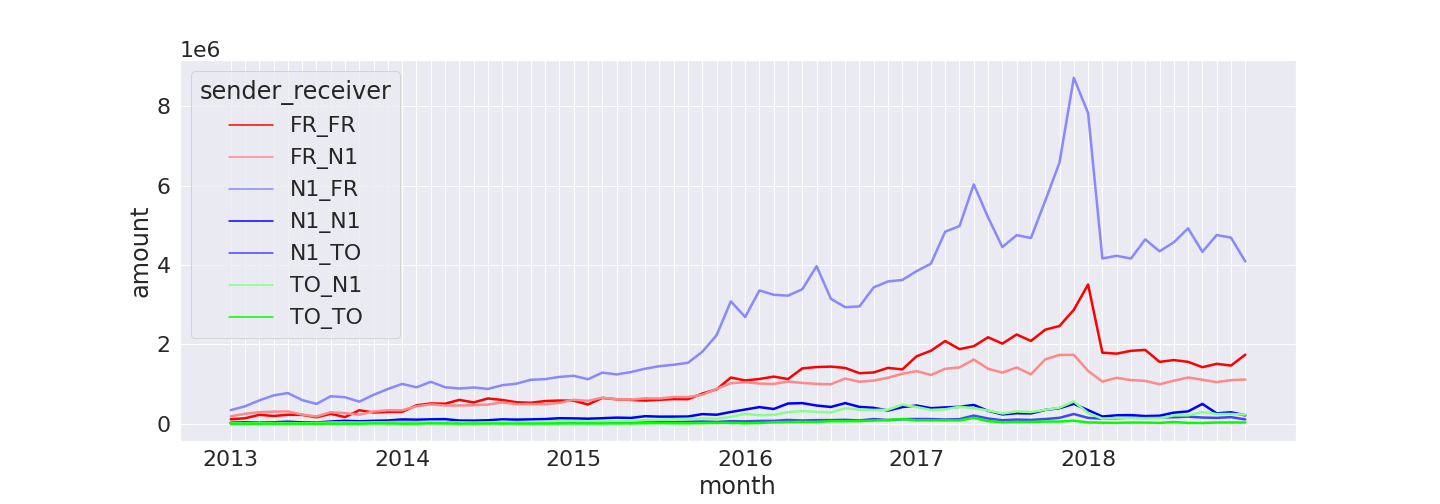}
    \includegraphics[width=0.75\textwidth,trim={0 30px 0 0},clip]{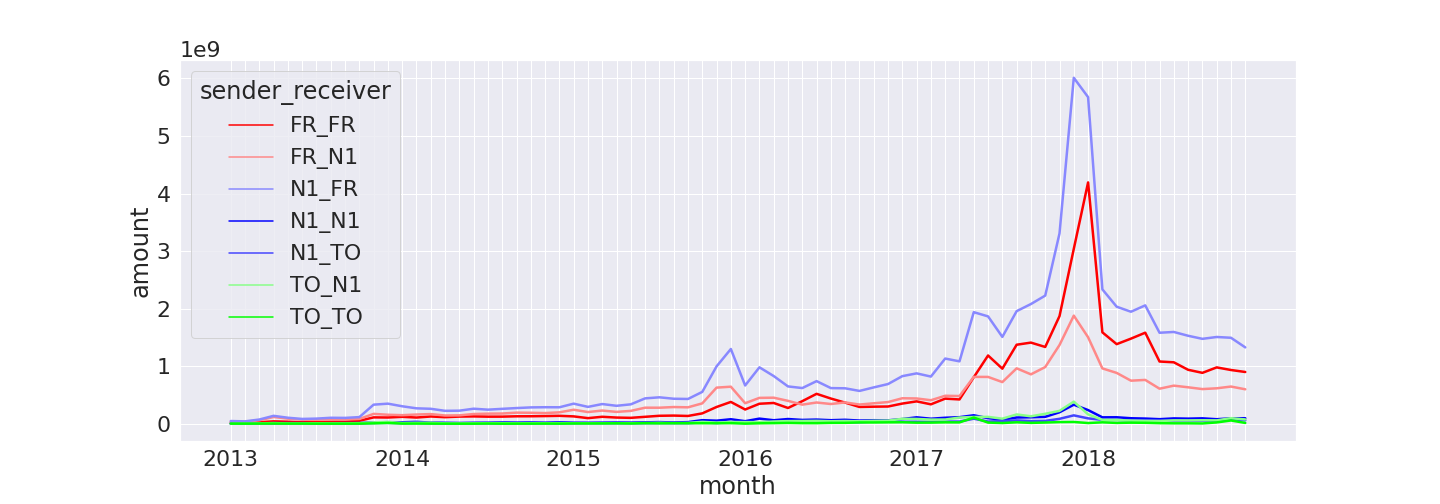}
    
\caption{Exchanges between categories. CAT1\_CAT2 in the legend identifies transactions from CAT1 to CAT2. Number of payments between categories (top). Amount of USD exchanged between categories (bottom).}
\label{img_usd_cat_to_cat}
\end{figure}

\noindent\textbf{Payments between categories.}
We turn our attention to the volumes exchanged between pairs of categories. Figure \ref{img_usd_cat_to_cat} shows the number of payments and the amount of USD exchanged between categories.
Most payments and USD amounts are transferred from N1 to FR entities, while FR to N1 also has a high volume, although in third position overall. We can thus confirm that transactions from N1 to FR entities are the most common in what we consider real economic exchanges between individuals and companies. 
Another common type of exchange occurs among FR entities (i.e., from one FR to another FR). This result may appear surprising at first, but it can potentially be explained by transactions initiated by exchange customers. Many individuals use exchange platforms (considered as FRs) similar to retail banks, where they can request a Bitcoin equivalent of bank transfers. In other words, they ask the exchange platform to make payments on their behalf to another person or company. The majority of FR-to-FR payments probably represent this kind of payments. However, it is worth further investigation since it could also involve exchanges between businesses (B2B) or artificial transactions resulting from complex fund management.

    

\subsection{Temporal analysis}
\begin{figure}[t]
\centering
\captionsetup[subfigure]{justification=Centering}

\begin{subfigure}[t]{0.40\textwidth}
    \includegraphics[width=\textwidth,trim={0 0 50px 0},clip]{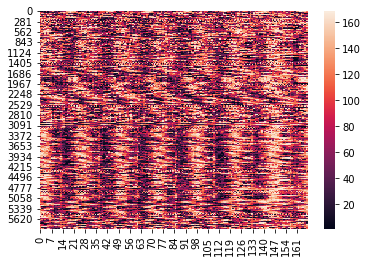}
    \caption{Normalized patterns,\\before alignment}
    \label{before_aligment_FILTERED}
\end{subfigure}
\begin{subfigure}[t]{0.45\textwidth}
    \includegraphics[width=\textwidth]{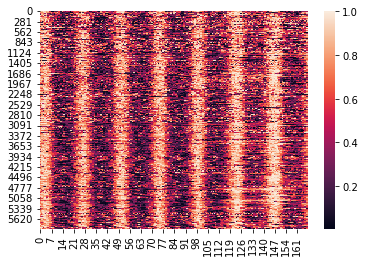}
    \caption{Normalized patterns,\\after alignment}
    \label{after_alignment}
\end{subfigure}

\caption{Normalized patterns of activities for all FR and years. Each row is a FR entity, each column corresponds to an hour of an average week. Dark colors correspond to low values, light colors to high values.}
\end{figure}

Since FR entities are supposed to have frequent interactions with clients, we can study their behavior in more detail from the rich information about payments they are involved in. This section analyzes the patterns of their weekly average behaviors.

\subsubsection{Hourly behavioral patterns.} We investigate temporal activities in an average week for entities identified as frequent receivers (FR). For each entity, we compute the payment volume for every weekly hour over a year with the following procedure.

\begin{enumerate}
    \item We define a vector $w$ of length $24 \times 7 = 168$, corresponding to the hours of an average week, i.e., $w[0]$ corresponds to Monday from 0 a.m. to 1 a.m., and $w[167]$ to Sunday from 11 p.m to 12 p.m.
    \item For each entity, we count the total number of payments received for each hour in a week according to Anywhere on Earth (AoE) time.
    \item Normalize the vector so that the sum of each entity's vector equals to 1.
\end{enumerate}

We remove some noisy data for FR entities with a likely non-human pattern of activity: those with more than 80 zeros in their mean weekly activity pattern, i.e., more than 80-hour slots without a single transaction during the year. The idea is that if an entity is indeed a company, it should receive transactions at nearly any time. Note that we use received payments, whose arrival time cannot be controlled by the receiver.
Figure \ref{before_aligment_FILTERED} shows that most FR (rows in the heatmap) have a regular pattern for each day of the week, thus following the pattern of typical human behavior.

\subsubsection{Temporal alignment.}

From Figure \ref{before_aligment_FILTERED}, it is evident that weekly patterns among entities are not aligned. This is expected due to the global nature of Bitcoin users residing in different time zones. Therefore, we assume that daily activity follows a similar pattern everywhere on average. For example, if the peak activity occurs at 2 p.m. in Japan, it would also occur at 2 p.m. in New York --- in the local time zones. This assumption is an approximation: each country has its unique characteristics, and some services might attract customers at different hours. Nevertheless, we believe it is reasonable for most entities, allowing for a deviation of up to two hours.

We propose finding an optimal alignment that minimizes the sum of absolute differences between weekly patterns. As we couldn't find an existing method in the literature, we developed a custom method (Algorithm~\ref{alg:align}) to solve the alignment task. We use the Mean Absolute Error (MAE) to the median of all patterns as objective, as it is less sensitive to large deviations than the Mean Square Error, to account for the presence of large values which follow a power-law distribution rather than a normal one.

\begin{algorithm}[htb]
\KwData{$W$: \textit{Weekly entities activity matrix} } 
$M \gets$ medianByCol$(W)$\;
$W^S \gets W$\;
$S \gets [0]*W$.nbRow\;
$previousTotalError \gets +\infty $ \;
$currentTotalError \gets W.$length$*2$\;
\While{previousTotalError $>$ currentTotalError}{
    $previousTotalError \gets currentTotalError$ \;
    $currentTotalError \gets 0 $\;
    \For{i \textbf{in} W.$nbRow$}{
        $w \gets W_i$
        $minError \gets +\infty $  \;
        $bestShift \gets NULL$ \;
        \For{shift \textbf{in} [0..23]}{
            $w^S \gets $shifted$(W_i,shift)$
            $error \gets 
            MAE(w^S,W_i)$ \;
            \If{error $<$ minError}{
                $minError \gets error$ \;
                $bestShift \gets shift$ \;
            }
            $W^S_i \gets w^S$

            $S[i] \gets bestShift $ \;
    currentTotalError $\gets$ currentTotalError + minError \;
    }
    }
}
\caption{Aligning process by minimizing the mean absolute error between normalized weekly patterns}\label{alg:align}
\end{algorithm}

Figure~\ref{after_alignment} displays the normalized activity patterns after the alignment process. The alignment algorithm yields two primary outcomes: (1) A vector $S$ that assigns a time zone shift to each entity, and (2) A matrix $W^S$ of weekly patterns where the goal is to align the patterns of all entities as closely as possible.

\subsubsection{Validation of the alignment.}
To assess the effectiveness of our alignment process, we carefully handpicked 10 entities from the WalletExplorer collection of known users. We specifically chose these entities because we could confidently assign them a country of usage ---not only domiciliation, thereby establishing a ground-truth time zone through manual research on their websites or historical data.
Table \ref{tab:timeZoneValid} presents the list of these entities along with their expected approximate time zone and the corresponding estimated time zone shift. In most cases, we can accurately identify the region of the world with a reasonable level of precision.


\begin{table}[t]
    \centering
    \begin{tabular}{c|c|c|c}
     \hline
     Entity & Expected Country & Approx. Time Zone & Avg. Estimated Shift \\ [0.5ex] 
     \hline\hline 
     MeXBT.com & Mexico & GMT-6 & -4.00 \\
     MercadoBitcoin.com.br & Brazil & GMT-3 & -2.12 \\
     FoxBit.com.br & Brazil & GMT-3 & -1.50 \\
     Paymium.com & France & GMT+2 & +1.80 \\
     SimpleCoin.com.cz & Czech Republic & GMT+2 & +2.00 \\
     BTC-e.com & Russia & GMT+3 & +3.00 \\
     Exchanging.ir & Iran & GMT+3:30 & +4.50 \\
     BX.in.th & Thailand & UTC+07 & +8.75 \\
     Huobi.com & China & GMT+08 & +9.67 \\
     CoinSpot.com.au & Australia & GMT+8/GMT+10 & +11.40 \\
     Bitfinex.com & Unknown & Unknown & -3.50 \\
     SilkRoadMarketPlace & Unknown & Unknown & -3.00 \\
     \hline
     \multicolumn{4}{c}{} \\[-1ex] 
    \end{tabular}
    \caption{Time zone of the 10 selected entities. We can observe that the estimated shift is close to the expected time zone. For the two entities at the bottom, for which we do not have \textit{a priori} knowledge of the location, the method assigns a time zone that we can interpret as being located in the Americas.}
    \label{tab:timeZoneValid}
\end{table}

\subsubsection{Estimation of Bitcoin's activity geographical distribution.}
\begin{figure}[ht!]
\centering
\captionsetup[subfigure]{justification=Centering}

\begin{subfigure}[t]{0.43\textwidth}
    \includegraphics[width=\textwidth,trim={0 20px 0 0},clip]{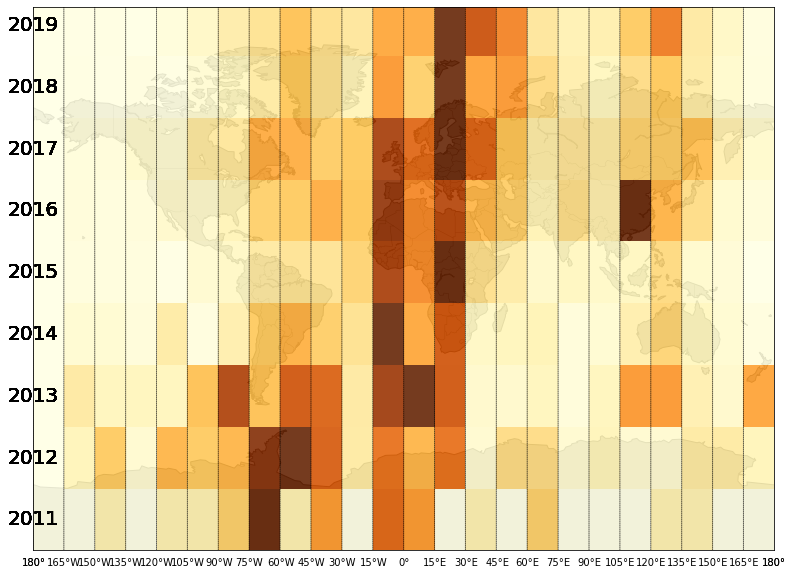}
    \caption{All FR, yearly normalization}
\end{subfigure}
\begin{subfigure}[t]{0.43\textwidth}
    \includegraphics[width=\textwidth,trim={0 20px 0 0},clip]{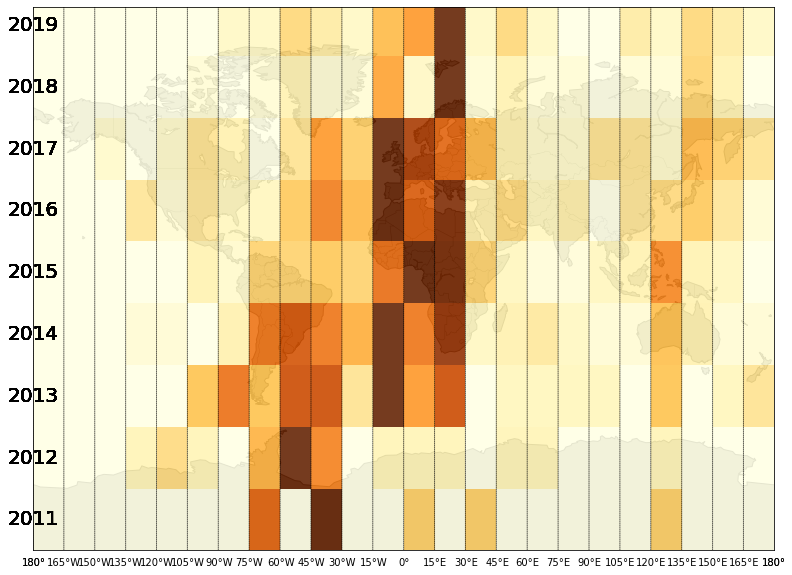}
    \caption{WalletExplorer only, yearly normalization}
\end{subfigure}

\begin{subfigure}[t]{0.43\textwidth}
    \includegraphics[width=\textwidth,trim={0 20px 0 0},clip]{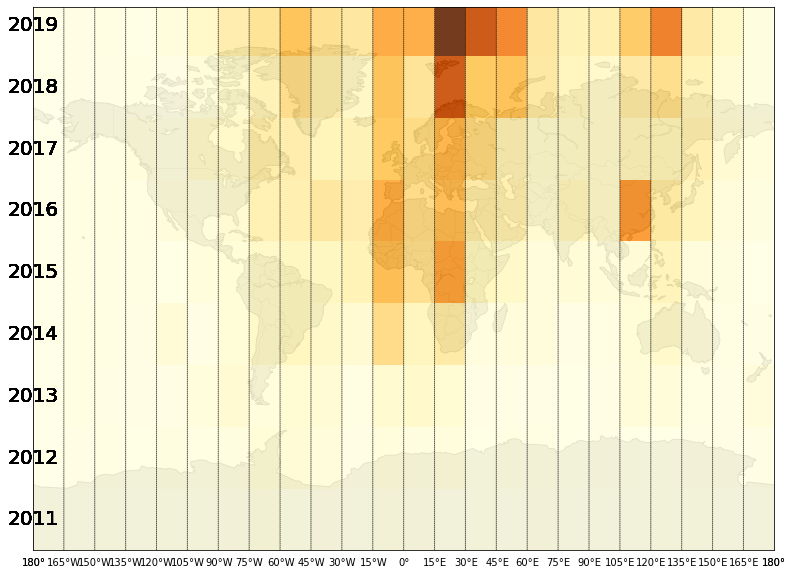}
    \caption{All FR, global normalization}
\end{subfigure}
\begin{subfigure}[t]{0.43\textwidth}
    \includegraphics[width=\textwidth,trim={0 20px 0 0},clip]{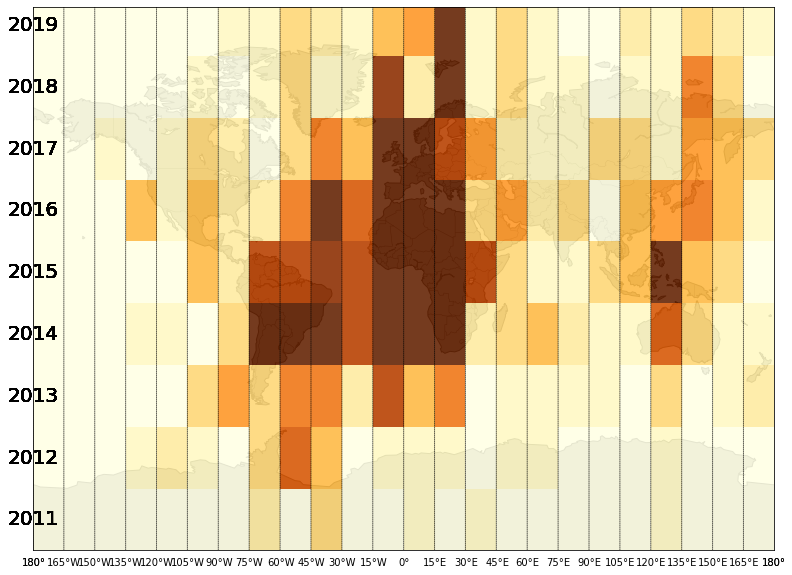}
    \caption{WalletExplorer only, global normalization}
\end{subfigure}
\caption{Number of FR entities identified by time zone and by year. Each horizontal line corresponds to a year, and each vertical line to a time zone. The continents' contours are provided for easier interpretation of the time zone information.}
\label{timeZones}
\end{figure}

The alignment process allows us to estimate the time zone of FR entities. Although it provides an estimation, and does not distinguish between countries with similar time zones, it enables us to estimate the geographical distribution of Bitcoin’s main players and track how this distribution evolves over time. Figures \ref{timeZones} (a and b) presents the number of entities for each time zone over the years, normalized by year(row), i.e., $c^{yearly}(y,h)=\frac{c(y,h)}{\sum_{i\in H}c(y,i)}$, where $H$ is the 24 possible hour shifts. We compare this with a subset composed only of known entities from the widely used WalletExplorer\cite{walletexplorer} dataset. Both datasets display similar patterns, with a little less activity in Asia for the WalletExplorer one. WalletExplorer thus has a reasonable geographical representativeness of FR entities. We also observe a shift in time from an activity initially concentrated in the Americas to a larger concentration in the Euro-African time zones.

Figures \ref{timeZones} (c and d) display the same data, but with a global normalization, i.e., $c^{global}(y,h)=\frac{c(y,h)}{\sum_{x \in Y,i \in H}c(x,i)}$, where $Y$ is the set of studied years. We observe a clear difference between the two. For all FR entities, we observe that the latest years concentrate most of the entities and that the historical importance of American entities nearly disappear compared with the later ones. East-European and African time zones seem to be the fastest growing and dominate the number of entities. Asia was particularly important for some years and regained importance at the end of our dataset.

On the other hand, we can observe that the WalletExplorer dataset has a completely different pattern. Most entities were active between 2014 and 2017, leading to a strongly biased view of the activity of the main players in the Bitcoin economy. This observation is coherent with the warning present on the WalletExplorer website, mentioning that \textit{“Name database is NOT updated (except some very rare cases) since 2016”}\footnote{\url{https://www.walletexplorer.com/info}}. Although this limitation is known, our work allows a quantitative estimation for the loss in representativeness, still frequently used (e.g., \cite{sun2022bitanalysis}). We can confirm this observation through the data: the fraction of all FR entities known in WalletExplorer declined from 30\% in 2014 to less than 3\% in 2019.


\section{Discussion}

Our work contributes to the understanding of the nature of Bitcoin real activity flow between different entity types.

We observed that a small number of commercial entities (FR) receive a massive amount of the payments, while customers (N1) make most of these payments. This is worrying for the decentralized nature of Bitcoin as cryptocurrency: most of the economic activity seems to go through central entities playing a role similar to the one of bank in the standard economic ecosystem.

In addition, we could locate entities geographically based on their activity. The time zone inference presented here should be viewed as an approximate representation rather than an unbiased reflection of reality. A surprising feature of the data is the minor importance of the American Continent in the latest years. A possible explanation can be that entities in North America in particular are global players, used all over the world. Hence, the algorithm may attribute them to an average value in the middle of the map. In future work, we could differentiate between local and global players by analyzing the amplitude of the weekly pattern. Indeed, a more international entity is likely to have a flatter temporal pattern compared to a national player with a more distinct one.

\section{Acknowledgment}
This work is partially supported by BITUNAM Project ANR-18-CE23-0004.

%
%

%
%
%
 \bibliographystyle{splncs04}
 \bibliography{mybibliography}

\begin{thebibliography}{10}
\providecommand{\url}[1]{\texttt{#1}}
\providecommand{\urlprefix}{URL }
\providecommand{\doi}[1]{https://doi.org/#1}

\bibitem{ante2022profiling}
Ante, L., Fiedler, F., Steinmetz, F., Fiedler, I.: Profiling turkish
  cryptocurrency owners. Available at SSRN 4248849  (2022)

\bibitem{ante2022individual}
Ante, L., Fiedler, I., Meduna, M.v., Steinmetz, F.: Individual cryptocurrency
  investors: Evidence from a population survey. International Journal of
  Innovation and Technology Management  \textbf{19}(04),  2250008 (2022)

\bibitem{aoyama2022cryptoasset}
Aoyama, H., Fujiwara, Y., Hidaka, Y., Ikeda, Y.: Cryptoasset networks: Flows
  and regular players in bitcoin and xrp. PloS one  \textbf{17}(8),  e0273068
  (2022)

\bibitem{dalal2021identifying}
Dalal, S., Wang, Z., Sabharwal, S.: Identifying ransomware actors in the
  bitcoin network. arXiv preprint arXiv:2108.13807  (2021)

\bibitem{di2018data}
Di~Francesco~Maesa, D., Marino, A., Ricci, L.: Data-driven analysis of bitcoin
  properties: exploiting the users graph. Journal of Data Science and Analytics
   \textbf{6},  63--80 (2018)

\bibitem{harrigan2016unreasonable}
Harrigan, M., Fretter, C.: The unreasonable effectiveness of address
  clustering. In: Ubiquitous intelligence \& computing. pp. 368--373. IEEE
  (2016)

\bibitem{hiramoto2020measuring}
Hiramoto, N., Tsuchiya, Y.: Measuring dark web marketplaces via bitcoin
  transactions: From birth to independence. Forensic Science International:
  Digital Investigation  \textbf{35},  301086 (2020)

\bibitem{walletexplorer}
Janda, A.: Walletexplorer.com: smart bitcoin block explorer

\bibitem{kondor2014rich}
Kondor, D., P{\'o}sfai, M., Csabai, I., Vattay, G.: Do the rich get richer? an
  empirical analysis of the bitcoin transaction network. PloS one
  \textbf{9}(2),  e86197 (2014)

\bibitem{liu2021characterizing}
Liu, X.F., Ren, H.H., Liu, S.H., Jiang, X.J.: Characterizing key agents in the
  cryptocurrency economy through blockchain transaction analysis. EPJ Data
  Science  \textbf{10}(1), ~21 (2021)

\bibitem{loporchio2023bitcoin}
Loporchio, M., Bernasconi, A., Di~Francesco~Maesa, D., Ricci, L.: Is bitcoin
  gathering dust? an analysis of low-amount bitcoin transactions. Applied
  Network Science  \textbf{8}(1),  1--28 (2023)

\bibitem{lorenz2020machine}
Lorenz, J., Silva, M.I., Apar{\'\i}cio, D., Ascens{\~a}o, J.T., Bizarro, P.:
  Machine learning methods to detect money laundering in the bitcoin blockchain
  in the presence of label scarcity. In: ACM International Conference on AI in
  Finance. pp.~1--8 (2020)

\bibitem{meiklejohn2013fistful}
Meiklejohn, S., Pomarole, M., Jordan, G., Levchenko, K., McCoy, D., Voelker,
  G.M., Savage, S.: A fistful of bitcoins: characterizing payments among men
  with no names. In: Internet measurement conference. pp. 127--140 (2013)

\bibitem{nakamoto2008bitcoin}
Nakamoto, S.: Bitcoin: A peer-to-peer electronic cash system. Decentralized
  business review p. 21260 (2008)

\bibitem{nemeczek2023insights}
Nemeczek, F., Weiss, D.: Insights on crypto investors from a german personal
  finance management app. Journal of Risk and Financial Management
  \textbf{16}(4), ~248 (2023)

\bibitem{sun2022bitanalysis}
Sun, Y., Xiong, H., Yiu, S.M., Lam, K.Y.: Bitanalysis: A visualization system
  for bitcoin wallet investigation. IEEE Transactions on Big Data  (2022)

\bibitem{tao2021complex}
Tao, B., Dai, H.N., Wu, J., Ho, I.W.H., Zheng, Z., Cheang, C.F.: Complex
  network analysis of the bitcoin transaction network. IEEE Transactions on
  Circuits and Systems II: Express Briefs  \textbf{69}(3),  1009--1013 (2021)

\bibitem{tubino2022towards}
Tubino, R.R., Robardet, C., Cazabet, R.: Towards a better identification of
  bitcoin actors by supervised learning. Data \& Knowledge Engineering
  \textbf{142},  102094 (2022)

\end{thebibliography}
%




 \end{document}